\documentclass[12pt,preprint]{aastex}

\shorttitle{RESONANT TEMPERATURE FLUCTUATIONS IN NEBULAE}
\shortauthors{Bautista \& Ahmed}
\begin{document}
\title{RESONANT TEMPERATURE FLUCTUATIONS IN NEBULAE IONIZED BY SHORT-PERIOD BINARY STARS}

\author{Manuel A. Bautista and Ehab E. Ahmed} 

\begin{abstract} 
A present prevailing open problem planetary nebulae research, and photoionized gaseous nebulae research at large,
is the systematic discrepancies in ionic abundances derived from
recombination and collisionally excited lines in many H~II regions and planetary nebulae.
Peimbert (1967) proposed that these discrepancies were due
to 'temperature fluctuations' in the plasma, but the amplitude of such fluctuations remain unexplained by standard 
phtoionization modeling.
In this {\it letter} we show that large amplitude temperature oscillations are expected to form in gaseous nebulae photoionized by
short-period binary stars. Such stars yield periodically varying ionizing radiation fields, which induce
periodic oscillations in the heating-minus-cooling function across the nebula. For flux oscillation periods of 
a few days any temperature perturbations
in the gas with frequencies similar to those of the ionizing source will undergo resonant amplification.
In this case, the rate of growth of the perturbations increases with the amplitude of the
variations of the ionizing flux and with decreasing nebular equilibrium temperature.
We also present a line ratios diagnostic plot  that combines [O~III] collisional lines and O~II
recombination lines for diagnosing equilibrium and fluctuation amplitude temperatures in gaseous nebulae.
When applying this diagnostic to the planetary nebula M~1--42 we find an equilibrium temperature of $\sim$6000~K
and a resonant temperature fluctuation amplitude ($T_{rtf}$) of $\sim$4000~K. This equilibrium
temperature is significantly lower than the temperature estimated  
when temperature perturbations are ignored.
\end{abstract}


\normalsize

\section{INTRODUCTION}

Arguably, the most intriguing question left unanswered in photoionization modeling in
astronomy pertains the origin of systematic discrepancies in ionic abundances derived from
recombination and collisionally excited lines in a large fraction of known H II regions and planetary nebulae (PNe).
Such differences in derived abundances can reach up to factors of 10 to 20 for C, N, O, and Ne in
some extreme PNe (Liu et al. 2000, 2001). Fifty years ago, Peimbert (1967) proposed that the discrepancies were due
to "temperature fluctuations" in the plasma, but the amplitude of such fluctuations needed to
reconcile the abundance determination are too large, in general, to be reproduced by standard
phtoionization modeling (Kingdon and Ferland 1995). The existence of temperature fluctuation of some sort has been
supported by modern spectra from high sensitivity, high spatial resolution instruments, notably the
Hubble Space Telescope (HST) among them. Liu et al. (2000, 2001) showed that the temperatures of PNe
determined by ratios of collisionally excited lines (e.g. [O III], [N II], [S III]) are typically larger
than the temperatures derived by fitting hydrogen recombination spectra (Te(Bac)). Moreover, the
abundance discrepancy factors from collisional and recombination lines from optical and UV
spectra of planetary nebulae and HII regions are correlated with the difference between Te([O III])
and Te(Bac). Further, temperature fluctuations from point-to-point determinations of electron
temperature have been obtained for several high surface brightness PNe and H II regions (Rubin et al. 2002, 2003;
Krabbe and Copetti 2002, 2005; O'Dell et al. 2003; Jones et al. 2016),
though the fluctuations are generally too small to be spatially resolved in detail.

At present, the idea that is receiving
the most attention in explaining the nature of temperature fluctuation is the hypothesis of nebular 
chemically inhomogenous. According to this hypothesis there would be in the nebula either pockets 
of cold very metal-rich plasma mixed with the gas (Liu et al. 2006) or an extended high metallicity gas embedded
in a less dense ambient gas with lower metallicity (Tsamis et al. 2003). Though, these models are difficult to justify
physically and lack workable models capable of making observationally testable predictions.

In recent years, it has been found that a large fraction of all
intermediate-mass stars, PNe progenitors are known to be in binary systems (Liu et al. 2006). Moreover,
there is now mounting observational 
evidence that binary stars play a significant role
in the PN ejection process. In perticular, it seems like all PNe with extreme abundance
discrepancy factors host short-period binary stars (Corradi et al. 2015; Wesson et al. 2016).
(Observational searches for close binaries demonstrate
that these are virtually all found in bipolar PNe (Tsamis et al. 2003). Though, because searches for binary stars
are mostly limited to near edge-on systems with significant photometric variability it remains
unproven whether all non-spherical PNe, which in fact are the large majority of PNe, are ejected
from binary stars. Surveys have found that between $\sim$10\% and 20\% of all PNe central stars are
close binary systems, with periods typically shorter than 3 days (Abt 1983). 

Here we show that the periodically varying photoionizing radiation field
of a short-period-binary-star will lead to resonant temperature fluctuations in
ionized cloud. The mechanisms for this process are described in the
next section. Further, Section 3 presents line ratios diagnostics to 
determine the equilibrium and resonant temperature fluctuation amplitude from
observed spectra.

\section{Resonant Temperature Fluctuations} 

Let us assume a stationary, inviscid and nonconducting medium in plane parallel symmetry. The equations of conservation of momentum and energy are
\begin{equation}
\frac{\partial u}{\partial t}= \frac{1}{m n_H}\frac{\partial P}{\partial x}
\end{equation}
and
\begin{equation}
\frac{\partial \epsilon}{\partial t} = \frac{P}{n_H}\frac{\partial u}{\partial x} +
L.
\end{equation}
Here, $u$ is the velocity, $n$ is the atomic density, $m$ is the average mass per particle,  $P$ is the pressure, $\epsilon$ is the internal energy,
$L$ is the heating-minus-cooling rate.
If the pressure is mostly due to gas pressure and the internal energy follows the perfect gas law
$$P = 2\times n_H T$$
and
$$\epsilon = 2\times (3/2) T,$$
where the temperature $T$ is given in units of energy. 
The factor of 2 in both relationships comes from the fact in a plasma where hydrogen is mostly ionized the number density of particles in the plasma is about twice the density of atoms.

Now, let us introduce a small temperature perturbation and assume that the density is constant through the cloud 
and in time, so that 
$$
T = T_0(x) + T_1(x,t).$$ 
Where $T_0$ is the steady state equilibrium temperature and $T_1$ is the time dependent perturbation.
Re-writting equations (1) and (2) in terms of temperature and combining them 
\begin{equation}
\frac{\partial^2 T_1(x,t)}{\partial t^2} = v_s^2\frac{\partial^2 T_1(x,t)}{\partial x^2}
+ \frac{\partial L}{\partial t}, 
\end{equation} 
where, $v_s^2=(2/3)(T_0/m)$ is the sound speed.

In steady state conditions $L$ is identically zero for $T=T_0$ and 
$\frac{\partial L}{\partial T} < 0.$
Hence, this term will damp any temperature fluctuations in the nebula. 

On the other hand, if the nebular ionizing source varies with time so will the heating and cooling rates. 
Let us assume a periodically ionizing source with angular frequency $\omega$, then heating-minus-cooling rates 
would oscillate around the equilibrium state as
\begin{equation}
\frac{dL}{dt}= L_1 sin(\omega t)
.
\end{equation}
In this case, the temperature fluctuation would have a solution
\begin{equation}
T_{rtf}(x,t)=\frac{L_1/3}{(kv_s)^2-\omega^2} sin(\omega t - kx) 
\end{equation}
These are resonant temperature fluctuations (RTF) powered by the variability of the ionizing source the propagating through the nebula.
The amplitude of such waves would resonate for $kv_s \approx \omega$. For short period binary stars with occultation periods of about 10~days the wavelength of the thermal waves is of the order of $10^{12}$~cm, which typically too small 
to be resolved observationally.

\begin{figure}
\rotatebox{0}{\resizebox{\hsize}{\hsize}
{\plotone{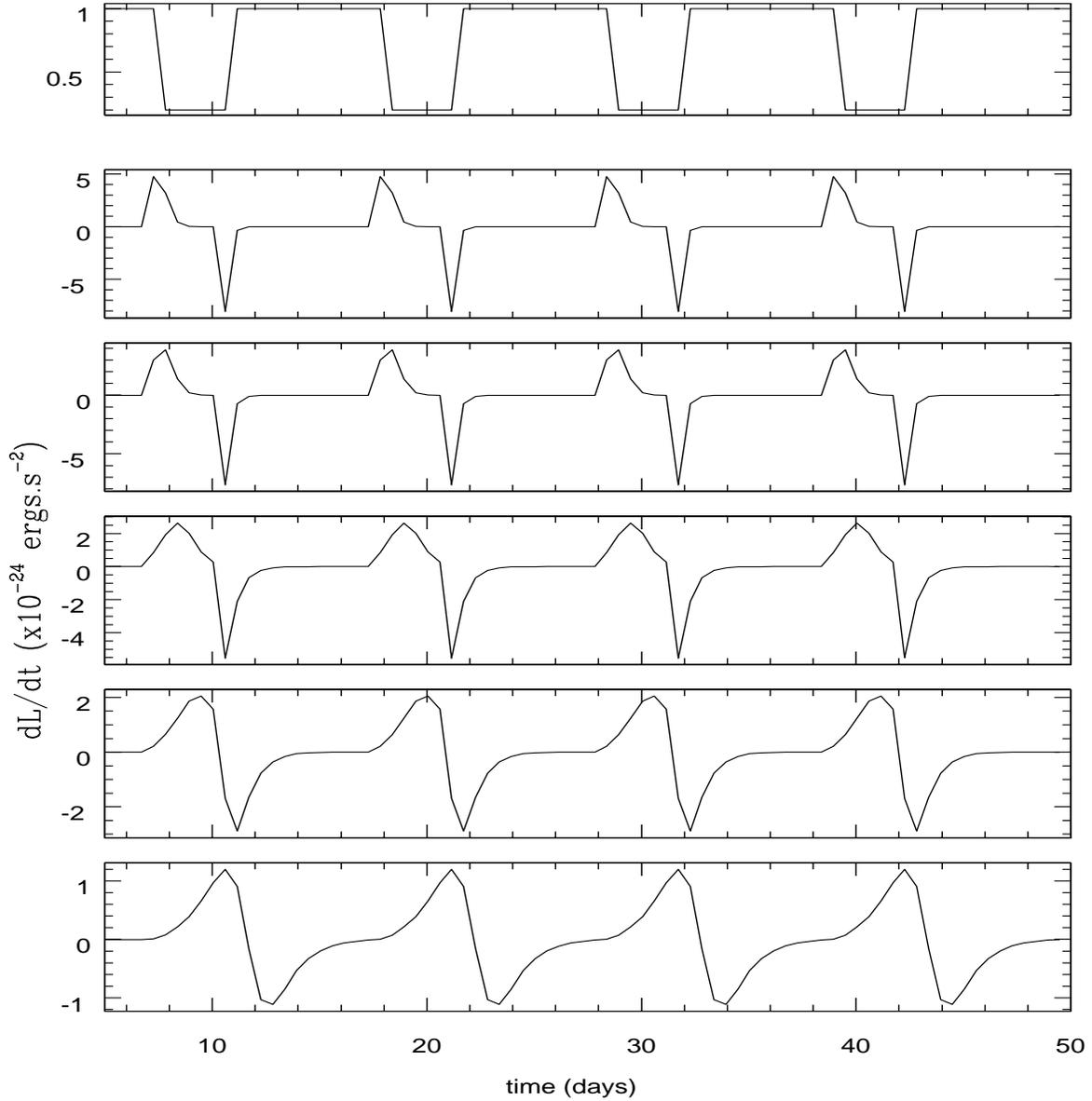}}}
\caption{Light curve of the ionizing source and first derivatives with respect to time of the heating-minus-cooling rates at different radii from the ionizing source. The 
different radii are $2.3\times 10^{14}$, $6.2\times 10^{14}$, $2.9\times 10^{15}$, $7.5\times 
10^{15}$, and $1.6\times 10^{16}$~cm. 
}
\label{hmcfunction}
\end{figure} 
 
Figure 1 shows the results of a simulation for a binary star with occultation period of 3 days. Here, the top panel
depicts the normalized light curve of the ionizing source.
We assume an spectral energy distribution given by a black body at 50,000~K and 
an ionization parameter at the spectral maximum $\log{(\psi)}=-1.0$.
The
circumstellar nebula has a constant density $n_H=10^4$~cm$^{-3}$ and 
a chemical mixture containing H, He, C, N, O, S, and Fe with solar abundances.
The five lower panels of Figure~1 
plot $\partial L/\partial t$ vs. time for various
radii within the nebula. The simulation was done with the time-dependent version of the photoionization
modeling code XSTAR (Ahmed 2017; Kallman and Bautista 1999).
The simulations show clear phase shifts in the heating-cooling wave for different radii in the nebula. This shift is
determined by the radiation propagation time (see Garc\'{\i}a et al 2013).

Clearly, the above treatment neglects additional damping terms. Because temperature waves travel at sound 
speeds, density fluctuations are also expected to occur. These may be the most important damping terms of
resonant temperature fluctuations.

Let us now consider the more general case of periodic perturbations to the system with the same frequency $\omega$ 
as the heating-minus-cooling function. This is,
$$T = T_0 + T_1(x,t) = T_0 + T_{rtf} \exp{[i(\omega t - k x)]}.$$
$$L = L_1(x,t) = L_1\exp{[i(\omega t - k_r x)]}$$
Notice here that while the radiation and gas properties are expected to oscillate with the same frequency 
they travel at different speeds, such that $\omega /k_r >> \omega/k$.

The linearized perturbation equations are
$$\frac{\partial n_1}{\partial t} = -n_0 \frac{\partial u_1}{\partial x} $$
$$u_0 \frac{\partial u_1}{\partial t} = \frac{1}{m} \left(\frac{T_0}{n_0}\frac{\partial n_1}{\partial x}
+ \frac{\partial T_1}{\partial x}\right) $$
$$3\frac{\partial T_1}{\partial t} = \frac{2 T_0}{m}\frac{\partial u_1}{\partial x} + L$$
This set of equations lead to the dispersion relation
\begin{equation}
T_{rtf}\times (k^2v_s^2-\omega^2)  = i \omega \frac{L_1}{3}. 
\end{equation}
For $T_{rtf}$ to be real the fluctuation's wave number must be complex as $k=k_r + i k_i$ and $k_r^2-k_i^2 = \omega^2
/ v_s^2$. Hence, the amplitude of the RTF amplitude is
\begin{equation}
T_{rtf} =\frac{\omega L_1}{6 v_s^2}\times \frac{1}{k_i \sqrt{(\omega/v_s)^2 + k_i^2}}.
\end{equation}

This equation shows that RTFs increase with 
the amplitude of ionizing flux variations and are inversely proportional to the equilibrium 
temperature of the nebula $T_0$ and to the binary occultation period.
A lower limit to $k_i$ is $\omega/v_r$, where $v_r$ is propagation velocity of radiation 
fronts, which is of the order of $0.1\times c$ (see Garc\'{\i}a et al. 2013). As per the 
results shown in Figure~2, $L_1 \sim 5\times 10^{-20}$~ergs.s$^{-1}$. Then, for $T_0=10^4$~K 
and a radiation fluctuation period of 1~day one finds $T_{rtf}\sim 3000$~K.

There are limits to the minimum length scale in which temperature fluctuation occur. The scale is set by 
the electron mean free path, within which electron temperatures can deviate from average values. As estimated by
Ferland et al. (2016), 
the electron mean free path is $\lambda_{e - e}\approx 10^{11}(10^3/n_e)(T(eV)/10~eV)^2$~cm. This means that
for sounds speeds $v_s\approx 10^6$~cm/s, the maximum binary occultation period for temperature fluctuations is
$T_{binary} \le 10^5 (10^3$~cm$^{-3}/n_e)$~s $\approx 1\times (10^3$~cm$^{-3}/n_e)$~days.
Furthermore, resonant temperature fluctuation in photoionized nebulae are expected when ionized by 
short-period binary systems, with occultation periods of the order of days.

\section{Diagnostics of equilibrium and RTF temperatures}

Figure~2 shows [O~III] and recombination O~II to [O~III] line ratios as a function of the average  electron temperature
for various fluctuation temperature amplitudes. The [O~III] spectrum was computed with atomic data from Mendoza and 
Bautista (2014) and the recombination lines are computed from effective recombination coefficients of Liu et al. (2001). 
Here the ratios are computed for a density $n_e=1200$~cm$^{-3}$. The observed ratios depicted in the figure
correspond to the planetary nebula M~1--42 (Liu et al. 2001), which exhibits extreme oxygen abundance discrepancies,
by about a factor of 20, 
as estimated from collisional and recombination lines. 
Our temperature diagnostics depicted in Figure~2 show that the nebula is consistent with an
average temperature $T_0\approx 6000$~K and fluctuations with amplitude $T_{rtf}\approx 4000$~K. 
Though, more detailed diagnostics must allow for uncertainties in the measured line ratios as well as uncertainties
in the atomic data and their propagation through the spectral models (see Bautista et al. 2013).
The large RTF temperature foud in this object is explained by the relative low equilibrium temperature.
The figure also shows that if temperature fluctuations are ignored the [O~III] temperature diagnostic would indicate 
a temperature of about 9000~K and discrepancy in O$^{2+}$ abundance determinations from collisional and recombination lines of about one order of magnitude, as found by Liu et al.

Our present model that explains nebular temperatures in terms of equilibrium  and 
RTF temperatures 
is consistent with the model proposed by Peimbert (1967) in that Peimbert's temperature fluctuation parameter 
\begin{equation}
t^2 \approx \frac{1}{2}\left(\frac{T_{rtf}}{T_0}\right)^2
\end{equation}

\begin{figure}
\rotatebox{0}{\resizebox{\hsize}{\hsize}
{\plotone{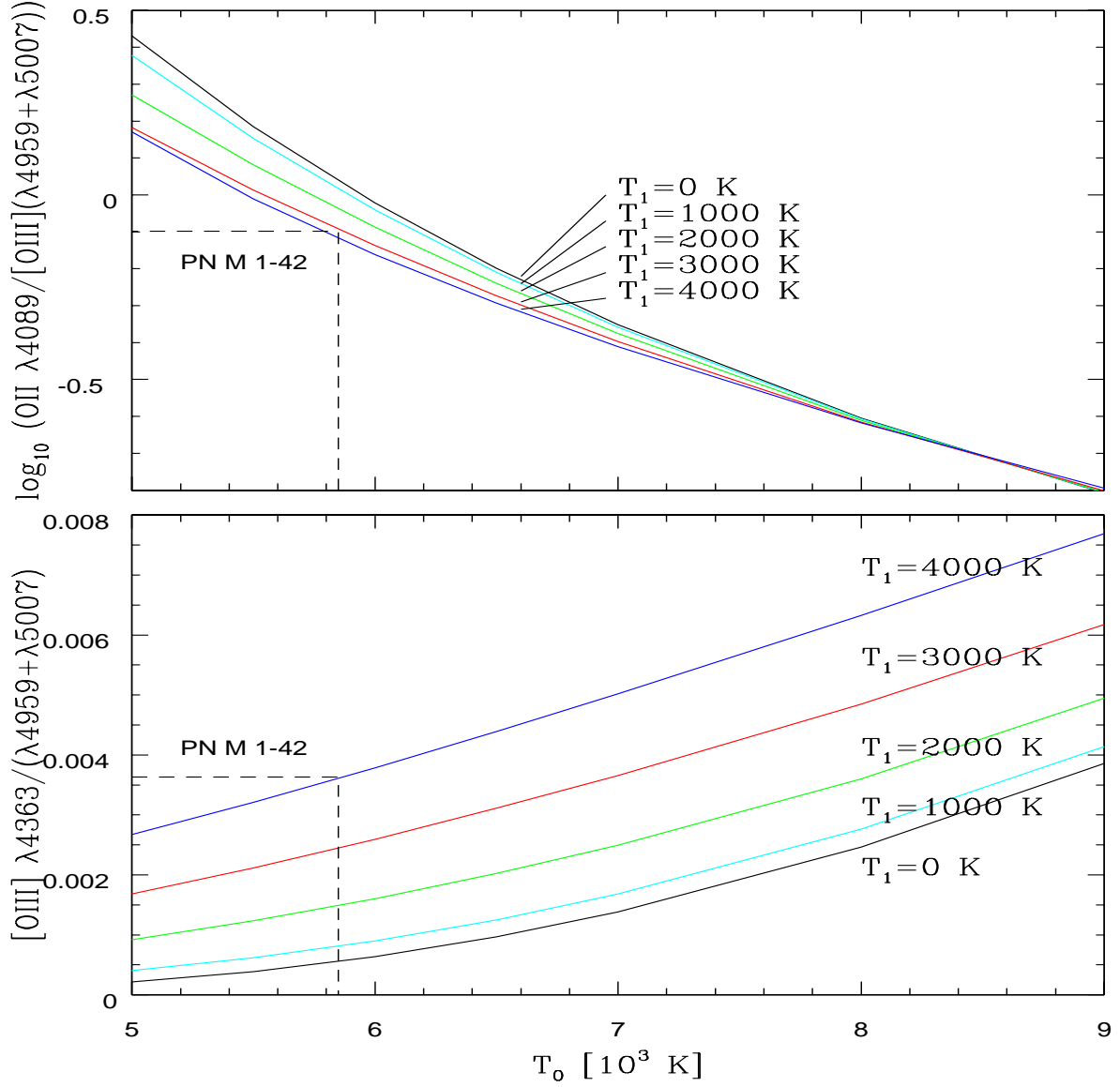}}}
\caption{Temperature diagnostic line ratios from collisionally excited [O III] line and recombination O II lines.
The line ratios are computed for an electron density $n_e = 1200$~cm$^{-3}$. Observed line ratios are for the 
planetary nebula M~1-42 (Liu et al. 2001).}
\label{lineratios} 
\end{figure}

\section{Conclusions}

We show that large amplitude (a few thousand kelvin) resonant amplitude fluctuations are expected to form in gaseous nebulae photoionized by
short-period-binary stars. Such systems yield a periodically varying ionizing radiation field, which induces
periodic oscillations in the heating-minus-cooling function across the nebula. As a result, temperature perturbations
in the gas with frequencies similar to those of the ionizing source will undergo resonant amplification.

Further study shows that the amplitude of the RTFs is proportional to the 
amplitude of the
variations of the ionizing flux
and inversely proportional
to the equilibrium temperature of the gas.
For typical conditions found in H~II regions and planetary nebulae the amplitude of RTFs
is of several thousand kelvin, in accord with current observational evidence of temperature fluctuations.

Further, we present a line ratios diagnostic plot  that combines [O~III] collisional lines and O~II
recombination lines for diagnosing equilibrium and fluctuation amplitude temperatures in gaseous nebulae. 
When applying this diagnostic to the planetary nebula M~1--42 we find and equilibrium temperature of $\sim$6000~K
and a temperature fluctuation amplitude of $\sim$4000~K. This equilibrium
temperature is significant lower than the 9000 - 10,000~K temperature diagnosed from [O~III] collisional
lines when temperature perturbations are ignored.
PN M~1--42 is known to exhibit extreme
temperature fluctuations, which is consistent with our prediction that resonant temperature perturbations should grow
faster with decreasing nebular equilibrium temperature.

Clearly, variations in the ionizing radiation from binary stars are more complicated than
a single mode plane wave. Though, these variations are periodic and as such they must be
well reproduced by linear combinations of plane waves. Thus, the present treatment can be
easily generalized to combinations of multiple modes of oscillation.

The temperature fluctuations described here are proper of nebulae ionized by sources 
that vary in intensity on a scale of few days. For sources varying on time scales of years,
comparable to the ionization and temperature equilibration time scales of the gas,
other type of temperature fluctuations will occur. These fluctuations are the result
of thermal fronts that propagate supersonically through the nebula. 
These fluctuations will be the subject of future publications.

\begin{acknowledgements}
This work was supported in part by the National Science Foundation (Award AST-1313265). 
\end{acknowledgements}



\end{document}